\documentclass[%
 aip,
 amsmath,amssymb,
 nofootinbib,
]{revtex4-1}

\usepackage{graphicx}
\usepackage{dcolumn}
\usepackage{bm}

\def\bfone{\relax{\rm 1\kern-.35em 1}}

\newcommand{\be}{\begin{equation}}
\newcommand{\ee}{\end{equation}}
\newcommand{\ben}{\begin{displaymath}}
\newcommand{\een}{\end{displaymath}}
\newcommand{\bea}{\begin{eqnarray}}
\newcommand{\eea}{\end{eqnarray}}

\newcommand{\bean}{\begin{eqnarray*}}
\newcommand{\eean}{\end{eqnarray*}}
\newcommand{\beqs}{\begin{eqnarray}}
\newcommand{\eeqs}{\end{eqnarray}}

\newcommand{\tg}{g}

\makeatletter \@addtoreset{equation}{section} \makeatother

\begin{document}
\title{Duality--symmetric actions for non--Abelian tensor fields}

\author{Igor Bandos}
 \affiliation{Department of Theoretical Physics, University of the Basque Country,\\
 UPV/EHU, P.O. Box 644, 48080 Bilbao, Spain \\
 and  IKERBASQUE, Basque Foundation for
Science, 48011, Bilbao, Spain}

\author{Henning Samtleben}%
\affiliation{
 Universit\'e de Lyon, Laboratoire de Physique, UMR 5672,
CNRS, \'Ecole Normale Sup\'erieure de Lyon, F-69364 Lyon
cedex 07, France
}%

\author{Dmitri Sorokin}
\affiliation{%
INFN, Sezione di Padova, via F. Marzolo 8, 35131 Padova, Italia 
}%


\begin{abstract}

We construct the duality--symmetric actions for a large class of six--dimensional models describing hierarchies of non--Abelian scalar, vector and tensor fields related to each other by first-order (self-)duality equations that follow from these actions. In particular, this construction provides a Lorentz invariant action for non-Abelian self-dual tensor fields.
The class of models includes the bosonic sectors of the $6d$ (1,0) superconformal models of interacting non--Abelian self--dual tensor, vector, and hypermultiplets.

\end{abstract}

\maketitle

\section{Introduction}
Understanding the detailed structure of the effective $6d$ theory of multiple $M5$--branes remains one of the
important long--standing issues of string/M--theory which, in particular, hampers the development of $AdS_7/CFT_6$
correspondence.
On general grounds this should be a (2,0) superconformal theory of non--Abelian chiral tensor supermultiplets
\cite{Witten:1995zh}. The theory does not have a free dimensionless parameter to make it weakly coupled and this casts doubts
on the very existence of its action. However, the action for a single M5--brane does exist
\cite{Pasti:1997gx,Bandos:1997ui,Aganagic:1997zq}\footnote{For an alternative construction based on the BLG model with the
gauge symmetry of volume preserving diffeomorphisms see \cite{Ho:2008nn,Ho:2008ve,Pasti:2009xc}. The equivalence of these models to \cite{Pasti:1997gx,Bandos:1997ui,Aganagic:1997zq} are still to be proved. } and produces the
M5--brane equations of motion \cite{Bandos:1997gm} first derived in \cite{Howe:1996yn} and analyzed in detail in \cite{Howe:1997fb,Howe:1997vn} using the superembedding techniques (see
\cite{Bandos:1995zw} and e.g. \cite{Bandos:1997qk,Howe:1997wf,Sorokin:1999jx,Bandos:2009xy} for a review and references). Various other aspects of the theory of $M$--branes are reviewed e.g. in \cite{Berman:2007bv,Simon:2011rw,Bagger:2012jb,Lambert:2012wr}. One may hope that
also for the multiple M5--branes an action may exist at least for a certain branch of the theory in which a
dimensionless coupling constant appears and makes perturbative Lagrangian description possible.

To make progress in the construction of the theory of multiple M5--branes one should first of all solve the
problem of consistently endowing the chiral tensor field with non--Abelian gauge structure, which itself is a
highly non--trivial problem. If one succeeds, one can then look for equations of motion and eventually for the
action.
Different ways of tackling these problems have been pursued. Several approaches have been aimed at rewriting and
re--interpreting the 6d theory (compactified on a circle) in terms of a 5d super--Yang--Mills theory
\cite{Lambert:2010wm,Lambert:2010iw,Douglas:2010iu,Singh:2011id,Lambert:2011gb,Lambert:2012qy,Ho:2011ni,Huang:2012tu,Chu:2012um,Chu:2012rk,Chu:2013hja,Bonetti:2012fn,Bonetti:2012st,Singh:2012qr,Kim:2012tr}.
In this way one gains a dimensional parameter (the radius of the compactification circle) which allows for
perturbative description.
 Other approaches use more sophisticated mathematical tools such as higher gauge
theories, twistor spaces and gerbes \cite{Kotov:2007nr,Baez:2010ya,Fiorenza:2012tb,Saemann:2012uq,Saemann:2012kj,Palmer:2012ya}. Each of the
approaches has its advantages, but also issues and limitations. However, one may hope that all these approaches
should be related to each other and can give us, from different perspectives, hints on what is a detailed
structure of the multiple M5--brane theory.

A more traditional field--theoretical approach based on the hierarchy of non--Abelian vector--tensor systems
\cite{deWit:2005hv} has been put forward in \cite{Samtleben:2011fj} (see also \cite{Chu:2011fd} for a particular
case) and further considered in \cite{Samtleben:2012mi,Akyol:2012cq,Samtleben:2012fb}. It aims at the construction
of superconformal models including non-Abelian tensor multiplets directly in 6-dimensional space-time showing that
a non--Abelian deformation of 6d chiral tensor fields is indeed possible upon further introducing higher rank
$p$-forms. Supersymmetrization of this construction and on-shell closure of the (1,0) supersymmetry algebra
produces unique equations of motion of the fields. It has further been shown that a sub-class of these models can
be promoted to have a pseudo-action in the sense that it reproduces all equations of motion except for the self
duality equations of the tensor fields. While such a pseudo-action may be considered as an efficient book-keeping
device for checking supersymmetry (and other symmetries) of the field equations, it does not provide a reliable
starting point for the quantization of the theory. In particular, one may wonder to which extent some of the
bothersome features of these models such as the apparent presence of ghosts in the scalar sector, and the
complicated vector field dynamics are an artifact of a pseudo-action.

The aim of this paper is to complete the construction of the actions for (1,0) superconformal theories initiated
in \cite{Samtleben:2011fj,Samtleben:2012mi,Samtleben:2012fb} by integrating the equations of motion to a fully--fledged self--dual
Lagrangian for the non--Abelian chiral 2--form fields and a duality--symmetric Lagrangian for vector gauge fields
and their 3--rank tensor duals. This construction yields a non--Abelian generalization of the covariant actions for $6d$ Abelian chiral tensor
fields \cite{Pasti:1996vs} and of their gauge--fixed counterparts
\cite{Henneaux:1987hz,Henneaux:1988gg,Schwarz:1993vs}.  This paper also generalizes and extends to $D=6$  the
results of \cite{Samtleben:2011eb} in which duality--symmetric but non--manifestly covariant actions for $D=4$ models with non--Abelian
twisted self--duality were constructed.

The paper is organised as follows: in Section~\ref{sec:hierarchy}, we review the general system of non-Abelian $p$-forms and their gauge transformations in six dimensions describing the
hierarchy of non--Abelian scalar, vector and tensor fields. The bosonic field equations for this system are given in Section~\ref{sec:10} as dictated by (1,0) superconformal symmetry.
They contain the non-Abelian self-duality equations for the tensor fields, while the vector field dynamics may be expressed in terms of a first-order duality equation relating their non-Abelian field strength to the field strength of the three-form gauge potentials.
The main part of the paper is Section~\ref{sec:action} in which we present an action that gives rise
to a general set of non-Abelian (self-)duality equations in six dimensions, including the bosonic sector of the (1,0) models as a special case. Space-time covariance is ensured by the presence of an auxiliary scalar field. We carefully analyse the Euler-Lagrange equations of the duality--symmetric action and show that their various bits cascade down to a combination of the first-order duality equations and derivatives thereof (which can be integrated as in the Abelian case). Together, the various parts of the equations of motion assemble into the full set of first-order (self-)duality equations. In Section~\ref{sec:example}, we work out some illustrative example and collect
our conclusions in Section~\ref{sec:conclusions}.


\section{Non-Abelian $p$--forms in six dimensions}
\label{sec:hierarchy}

In this section we will briefly review the system of non-Abelian $p$-forms ($p=1, \dots, 4)$
and their gauge transformations in six dimensions describing the
hierarchy of non--Abelian scalar, vector and tensor fields.
For more details the
reader is referred to \cite{Samtleben:2011fj,Samtleben:2012mi,Samtleben:2012fb}.

The tensor hierarchy is formed by the $p$-forms $(A^r_1,B_2^I,C_{3\,r},C_{4\,\alpha})$ or in component notation $(
A_\mu^r, B_{\mu\nu}^I, C_{\mu\nu\rho\,r},C_{\mu\nu\rho\lambda\,\alpha})$, with six-dimensional space-time indices
$\mu, \nu, \dots$. They couple to the scalars $( Y^{ij\,r},\phi^I )$, where $Y^{ij\,r}$ are part of the off--shell
vector multiplets, while $\phi^I$ together with $B_2^I$ and their fermionic partners form the chiral tensor
supermultiplet. Later we will also add the non--Abelian hypermultiplets. To avoid the proliferation of indices, we
will work with differential forms on which the external derivative will act from the right. In what follows we
will be only interested in a subclass of the models which have a Lagrangian description. This requires the
introduction of an (indefinite) constant metric $\eta^{IJ}$ and its inverse $\eta_{IJ}$
$(\eta^{IJ}\eta_{JK}=\delta^I_K)$ which raise, lower and contract the indices $I,J,K$.

The non--Abelian field strengths of the $p$-form gauge potentials are given by
\begin{eqnarray}\label{cF:=}
{\cal F}^r&:=&{1\over 2} dx^\nu\wedge dx^\mu \,{\cal F}_{\mu\nu}^r=  dA^r +{1\over 2} f_{st}{}^r A^s\wedge A^t
+g^r_IB_2^I \; , \qquad \\ \label{cH3:=} {\cal H}_3^I&:=&{1\over 3!} dx^\rho\wedge dx^\nu\wedge dx^\mu\, {\cal
H}_{\mu\nu\rho}^I= DB_2^I + d_{st}^IA^s\wedge dA^t  + {1\over 3} f_{pq}{}^s d_{rs}^I A^r\wedge A^p\wedge A^q
+g^{Ir} C_{3\, r}  =  \nonumber \\ &&   = dB_2^I + d_{st}^IA^s\wedge \left({\cal F}^t+ g^t_JB_2^J\right)   -
{1\over 6} f_{pq}{}^s d_{rs}^I A^r\wedge A^p\wedge A^q +g^{Ir} \left( C_{3\, r} - 2d_{Jrs} B_2^J\wedge A^s\right),
\nonumber \\ &&
 \qquad  \\ \label{cH4:=} {\cal H}_{4r}&:=&{1\over 4!}
dx^\sigma\wedge dx^\rho\wedge dx^\nu\wedge dx^\mu \,{\cal H}_{\mu\nu\rho\sigma \, r}=  DC_{3r} -2 B_2^J \wedge
\left(dA^s +{1\over 2} f_{pq}{}^s A^p\wedge A^q \right)d_{Jrs}  - \nonumber \\ && - B_2^J \wedge B_2^Id_{Irs}
g^s_J  + {1\over 3} dA^s\wedge A^t \wedge A^u d_{Js[t}d^J_{u]r} +k_{r}^{\alpha} C_{4 {\alpha}}  + \propto A\wedge
A\wedge A \wedge A\; .
\end{eqnarray}
They are constructed with the use of the antisymmetric `structure constants' $f_{st}{}^r=f_{[st]}{}^r$, the constant
tensors $d^I_{rs}=d^I_{(rs)}$ inducing Chern-Simons couplings, and the constant tensors $g^{Ir}$ and $k^\alpha_r$
that induce St\"uckelberg-type couplings among forms of different degree.  These tensors satisfy certain algebraic
relations which we have collected in appendix~\ref{app:con}. Notably, they satisfy the orthogonality relations
\bea g^{Ir}g_{I}^s:= g^{Ir} \eta_{IJ}g^{Js}=0\; , \qquad g^{Ir}k_{r}^{\alpha}=0 \;.\eea

The covariant derivatives $D$ are defined as follows
\begin{eqnarray}\label{DFr=d+}
&& D{\cal F}^r:=d{\cal F}^r + {\cal F}^t\wedge A^s X_{st}{}^r  =d{\cal F}^r - {\cal F}^t\wedge A^s f_{st}{}^r +
{\cal F}^t\wedge A^s d_{st}^Ig_I{}^r \; , \qquad
\\ \label{DcHI=d+} && D{\cal H}^I_3:=d{\cal H}^I_3 + {\cal H}^J_3\wedge A^s X_{sJ}{}^I
=d{\cal H}^I_3 + 2{\cal H}^J_3\wedge A^s d_{st}^Ig_J{}^t  - 2\,{\cal H}^J_3\wedge A^r g^{Is}d_{Jsr} \; , \qquad\\
&&D{\mathcal H_{4\,r}}:=d{\mathcal H_{4\,r}}- \mathcal H_{4\,t}\wedge A^s X_{sr}{}^t\,.\label{DH4}
\end{eqnarray}
Let us also note that the algebraic constraints among the constant tensors parametrizing the gauge system
in particular imply that the gauge generators appearing in these covariant derivatives are
related to the St\"uckelberg coupling $k_r^\alpha$ via
\bea \label{X=kc}
X_{rs}{}^t\; &\equiv& d^I_{rs}\,g_I^t-f_{rs}{}^t=-k_r^\alpha \,c_{\alpha\,s}^t\;,
\nonumber\\
X_{r\,IJ} &\equiv & 4 g_{[I}^sd_{J]\,rs}   = 2\,k_r^\alpha\,c_{\alpha\,IJ}
\;,
\eea
with tensors $c_{\alpha\,s}^t$ and $c_{\alpha\,IJ}$ whose role will be clarified below.
{}From \eqref{cF:=}--\eqref{DH4} one gets the Bianchi identities
\begin{eqnarray}\label{DcF=}
&& D{\cal F}^r= g_I^r {\cal H}_3^I \; , \qquad \\ \label{DcHI=} && D{\cal H}_3^I=
 d_{st}^I {\cal F}^s\wedge {\cal F}^t+g^{Ir} {\cal H}_{4r}
\; , \qquad \\ \label{DcH4=} && D{\cal H}_{4r} = - 2\,{\cal H}_3^I \wedge {\cal F}^sd_{Isr}+ k_r^\alpha {\cal
H}_{5\alpha}\; . \qquad
\end{eqnarray}
Eq.~\eqref{DcH4=} defines the 5--form field strength $\mathcal H_{5\alpha}=DC_{4\alpha}+\cdots$ (at least under projection with $k_r^\alpha$). We will not need
its explicit form in our construction. We only notice that $\mathcal H_{5\alpha}$ contains the  tensors $c_{\alpha\,s}^t$ and $c_{\alpha\,IJ}$ which enter eqs. (\ref{X=kc}), so that its Binachi identites read  \cite{Samtleben:2011fj}
\begin{eqnarray}\label{DcH5=}
D{\cal H}_{5\alpha} = -c_{\alpha\,IJ}  {\cal H}_3^I \wedge  {\cal H}_3^J -  c_{\alpha\,s}^r {\cal F}^s \wedge{\cal H}_{4r}+ \ldots  \; . \qquad
\end{eqnarray}
Actually, also neither explicit form of $\mathcal H_{4r} $ nor $\mathcal
H^I_{3}$ is needed for our calculations. The expressions for the general variation of the covariant field
strengths  \eqref{cF:=}--\eqref{cH4:=} can be reproduced formally from the Bianchi identities. These are
 \bea
\delta {\cal F}^r &=&  D \delta A^r + g^r_I\,\Delta B_{2}^I \;,\nonumber\\[.5ex] \delta {\cal H}_{3}^I &=&  D
\Delta B_{2}^I +2\, d^I_{rs} \, {\cal F}^r \wedge \delta A_{}^s + g^{Ir}\,\Delta C_{3\,r} \;,\nonumber\\[.5ex]
\delta {\cal H}_{4\,r} &=& D\Delta C_{3\,r} - 2d_{Irs}\,{\cal F}^s \wedge \Delta B_{2}^I -  2d_{Irs}\,{\cal
H}_{3}^I\wedge\delta A^s + k^\alpha_r\,\Delta C_{4\alpha} \;,\label{Delta2}
\eea
where we have introduced the
compact notation
\bea \Delta B^I_{2} &\equiv& \delta B^I_{2} + d^I_{rs}\,A^r \wedge \delta A^s \;,\nonumber\\
\Delta C_{3\,r} &\equiv& \delta C_{3\, r} -\,2 d_{Irs}\,B_{2}^I \wedge \delta A^s -{1\over 3} d_{Irs}\,
d^I_{pq}\,A^s \wedge A^p \wedge \delta A^q \;,\nonumber\\ k^\alpha_r \Delta C_{4\alpha}&\equiv & k^\alpha_r\delta
C_{4\alpha}+ \cdots \label{Delta1} \eea

The non--Abelian gauge transformations with $(p-1)$--form parameters
$(\Lambda^r,\Lambda^I_{1},\Lambda_{2\,r},\Lambda_{3\alpha})$ are given by \bea \label{gaugesym} \delta A^r &=& D \Lambda^r - g^r_I \Lambda_1^I \;,\nonumber\\[.5ex] \Delta B_{2}^I &=&
D\Lambda_1^I - 2d^I_{rs} \Lambda^r {\cal F}^s - \tg^{Ir} \Lambda_{2\,r} \;,\nonumber\\[.5ex] \Delta C_{3\, r} &=& D
\Lambda_{2\,r} +\, 2d_{Irs}\,{\cal F}^s \wedge\Lambda_{1}^I + 2d_{Irs}\,{\cal H}_{3}^I \,\Lambda^s-k^\alpha_r
\Lambda_{3\alpha} \;,
\\ [.5ex] k^\alpha_r\Delta C_{4\alpha} &=& k^\alpha_r D\Lambda_{3\alpha}-4X_{r\,IJ} \,\mathcal
H^I_3\wedge\Lambda^J_1
-X_{rs}{}^t\left( {\cal F}^s\wedge  \Lambda_{2t} +  \Lambda^s \,{\cal H}_{4t} \right)
\;.\label{gaugesym4} \eea
Under these transformations, the field strengths (\ref{cF:=}) transform covariantly
as $\delta {\cal F}^r= - {\cal F}^t \Lambda^s X_{st}{}^r$, $\delta {\cal H}_{3}^I= - {\cal H}_{3}^J \Lambda^s X_{sJ}{}^I$, etc.
Notice in particular, that the left-- and the right--hand--side of (\ref{gaugesym4}) vanish when contracted with $g^{Kr}$ in virtue of the identities \eqref{conaction}. For completeness, we also note that the connection in (\ref{gaugesym4}) is given by $k_r^\alpha\,A^s  X_{s\,\alpha}{}^\beta \equiv
A^s\,X_{rs}{}^t k_t^\beta$.
Consequently, also the 4-form field
strengths transform covariantly as
\bea
\delta {\cal H}_{4r}&=& 2\Lambda^s d^I_{rs}g_I^t\,{\cal H}_{4t} + {\cal F}^s\wedge X_{rs}{}^t \Lambda_{2t}
+ k_r^\alpha c_{\alpha\,s}^t \left(  {\cal F}^s\wedge  \Lambda_{2t} +  \Lambda^s \,{\cal H}_{4t}  \right)
\nonumber\\
&=&
 \Lambda^s X_{sr}{}^t\,{\cal H}_{4t}
 \;.
\eea

\section{Bosonic part of the $(1,0)$ superconformal field equations}
\label{sec:10}

So far, we have introduced the non-Abelian system of $p$-forms in six dimensions on a purely
kinematical level. Its supersymmetric dynamics may be deduced from closure of the
(1,0) supersymmetry algebra \cite{Samtleben:2011fj}. In particular, this fixes the couplings of
the $p$-forms to the scalar fields $\phi^I$ and $Y^{ij}$ completing the $(1,0)$ vector and tensor
multiplets, respectively.

In absence of hypermultiplets, and when all the fermions are set to zero, the resulting
bosonic field equations are
\bea \mathcal H^I_{3}+*\mathcal H^I_{3}&=&0 \;, \label{HH}\\[1ex]  D*D\phi_I-2d_{Irs}{\cal F}^r \wedge  *{\cal
F}^s-d^6x\,(2d_{Irs}Y^{ij\,r}Y_{ij}^s+3g^r_{(J}g^s_{K}d_{I)rs}\phi^J\phi^K)&=&0\;,
\label{eom_tensor} \eea
for the tensor multiplets and
 \bea \label{eom_scalar} d_{Irs}\,
Y^s_{ij} \,\phi^I  &=& 0 \;, \\[1ex]
 2d_{Irs}\phi^I\ast \mathcal F^s+\mathcal H_{4r}&=&0 \label{eom_vector}\;,
\eea for the vector multiplets.
Eq. \eqref{eom_scalar} reflects the auxiliary nature of the fields $Y^{ij\,r}$.
Eq. \eqref{HH} tells us that the 3--form field strength is self--dual and
eq. \eqref{eom_vector} is the first-order duality equation that relates the vector
field strengths to the field strengths of the three-form tensors. Its derivative together with the Bianchi
identities~(\ref{DcH4=}) yields the standard second-order Yang-Mills equation for the vector fields. In turn, the
four-form tensors are related by their field strength
to the scalar fields of the theory by means of the duality equation  \be\label{H5}
k^\alpha_r*\mathcal H_{5\alpha}=
\frac12\,{\cal J}_r \equiv
\frac12\,X_{r\,IJ}\, \phi^ID\phi^J\,, \ee
with the scalar matter current ${\cal J}_r$.
In presence of hypermultiplets, the
r.h.s.\ of this duality equation receives an additional contribution from the hyper scalar
current~\cite{Samtleben:2012fb}.

In the next section we will construct an action that reproduces the first-order (self-)duality
equations (\ref{HH}), (\ref{eom_vector}), (\ref{H5}) by extending the construction of \cite{Pasti:1996vs} to the non-Abelian case.

\section{The action}
\label{sec:action}

In this section we will present an action from which the field equations of the previous section are derived. In
particular, this includes an action for the non--Abelian chiral gauge field $B^I_2$.
More generally, we will construct an action which reproduces the general set of six-dimensional
non-Abelian (self-)duality equations for the $p$-forms
\bea
\mathcal H^I_{3}+*\mathcal H^I_{3}&=&0 \;, \nonumber\\[1ex]
\mathcal H_{4r} + {\cal M}_{rs} \ast \mathcal F^s&=&0\;,\nonumber\\[1ex]
2\,k^\alpha_r*\mathcal H_{5\alpha}-{\cal J}_r &=& 0\;.
\label{general-selfdual}
\eea
The particular choice of
\bea {\cal M}_{rs} = 2\, \phi^I\,d_{I\,rs}\;,\qquad {\cal J}_r =
X_{r\,IJ}\,  \phi^I D \phi^J \;, \label{MJ10}
\eea
for the vector kinetic matrix and the scalar current corresponds to the bosonic sector of the (1,0) superconformal models
 discussed in Section~\ref{sec:10} above, but our results apply to any six--dimensional system of the form (\ref{general-selfdual}). In particular, they include the coupling of the vector and tensor multiplets to the (1,0) hypermultiplets considered in \cite{Samtleben:2012fb}.

We will proceed in two steps. First, in Section~\ref{subsec:ct} we construct an action that gives rise to the non--Abelian self--duality equation (\ref{HH}) for the tensor fields together with the standard second-order field equations for the remaining fields. It is of the form
 \be\label{miniL} \mathcal S=
 \int_{\mathcal M^6} {\cal L}=
\int_{\mathcal M^6}({\cal L}^{\rm scal}+{\cal L}^{\rm vec} +{\cal L}^{\rm top}+{\cal L}^{\mathcal{HH}})\,. \ee
The first three terms in \eqref{miniL} which include kinetic terms of the scalars and the vector gauge field have
been constructed in \cite{Samtleben:2011fj}. The last term ${\cal L}^{\mathcal{HH}}$ is the Lagrangian for the
non--Abelian chiral gauge field $B^I_2$ whose construction is one of the main results of this paper.
In the second step,
in Section~\ref{subsec:vt}, we generalize this action to a duality--symmetric action that also treats vector fields and three-form
gauge potentials on the same footing and produces their first--order duality equation (\ref{eom_vector}) among
the proper field equations.
This is achieved by extending (\ref{miniL}) to
\bea
\label{Lext}
\mathcal S_{\rm ext}=
 \int_{\mathcal M^6} {\cal L}_{\rm ext} &\equiv&
 \int_{\mathcal M^6} \left({\cal L}^{\rm scal}+{\cal L}^{\rm vec}
+{\cal L}^{\rm top}+{\cal L}^{\mathcal{HH}} +{\cal L}^{{\cal H}_4/{\cal F}}\right)
\;,
\eea
with the new term ${\cal L}^{{\cal H}_4/{\cal F}}$ carrying the field strength of the three-form
gauge potentials.

In the differential form notation the first term in the actions \eqref{miniL}, \eqref{Lext}
has the following generic form
\be\label{Lphi} \mathcal
L^{\rm scal}=-\frac{1}{2} D\phi^I \wedge *D\phi^J\eta_{IJ}-V_{\rm scal}\,d^6x\;,
\ee
with covariant derivatives $D$ and where $d^6x$ stands for the 6--form   $dx^{\mu_1}\wedge ... \wedge
dx^{\mu_6}=\varepsilon^{\mu_1\ldots \mu_6}d^6x$. The scalar potential $V_{\rm scal}$ is a priori arbitrary. In the case of the $(1,0)$ superconformal models of \cite{Samtleben:2011fj} it takes the following form
\bea
V_{\rm scal} &\equiv&
-d_{Irs}(2\phi^IY^{ij\,r}Y_{ij}^s +
g^r_Jg^s_K\phi^I\phi^J\phi^K)\;,
\label{potential}\eea
with additional contributions in the presence of hypermultiplets~\cite{Samtleben:2012fb}.
The kinetic term for the vector fields in \eqref{miniL}, \eqref{Lext}
is of the standard form
\begin{eqnarray}\label{LcF*cF=}
  {\cal L}^{\rm vec}= {\cal M}_{rs} \, {\cal F}^r\wedge *{\cal F}^s
     \; \qquad
\end{eqnarray}
where the matrix ${\cal M}_{rs}$ is constructed from the scalars. In the case of the $(1,0)$ superconformal models it is defined by (\ref{MJ10}) in terms of the tensor multiplet scalars.

The presence of the topological term ${\cal L}^{\rm top}$ in the action (as in the other cases of this kind) is due to
the presence of Chern--Simons--like terms in the covariant field strengths \eqref{cF:=}--\eqref{cH4:=}. It is
constructed as follows. The $6d$ space--time $\mathcal M^6$ is formally extended to a $7d$ manifold $\mathcal M^7$
assuming $\mathcal M^6$ to be the boundary of $\mathcal M^7$ ($\mathcal M^6=\partial \mathcal M^7$). Then using
the field strengths \eqref{cF:=}--\eqref{cH4:=} one constructs the $7d$ form
\begin{equation}\label{dLtop=}
d{\cal L}^{\rm top}  := - 2d_{Ist} {\cal F}^s \wedge {\cal F}^t \wedge {\cal H}_{3}^I  + {\cal H}_{3}^I \wedge D{\cal
H}_{3}^J\eta_{IJ} =
 - d_{Ist} {\cal F}^s \wedge {\cal F}^t \wedge {\cal H}_{3}^I  +
{\cal H}_{3}^I \wedge g_I^r {\cal H}_{4r}\,,
\end{equation}
which is (identically) closed $dd{\cal L}^{\rm top}\equiv 0$, as can easily be checked using the Bianchi identities
(\ref{DcF=}), (\ref{DcHI=}) and (\ref{DcH4=}).  Then the topological action is
\be
\label{top} {\cal S}^{\rm top}
= \int_{\mathcal M^7} ({\cal H}_{3}^I \wedge g_I^r {\cal H}_{4r}- d_{Ist} {\cal F}^s \wedge {\cal F}^t \wedge {\cal H}_{3}^I
) =\int_{\mathcal M^6=\partial \mathcal M^7}{\cal L}^{\rm top}\,.
\ee
For performing the variation
of the action we do not need the explicit form of ${\cal L}^{\rm top}$, since $\delta {\cal L}^{\rm top} =
i_\delta (d{\cal L}^{\rm top})+ d(i_\delta {\cal L}^{\rm top})$ and the second term does not contribute to the
integral when the $6d$ space is assumed to have no boundaries \footnote{$i_\delta$ is the contraction operation with the variation $\delta$ considered as a vector field, so that $i_\delta d=\delta$, $i_\delta dA^r= \delta A^r$ etc. In our conventions this operation acts from the right, \emph{e.g.}
 $i_\delta (d\phi^I\wedge d\phi^J)= d\phi^I\, \delta \phi^J- \delta \phi^I\, d\phi^J$.
 }. Actually we also use this property for other Lagrangian forms and omit total derivative terms in their variation.

The following construction applies to arbitrary
scalar and vector couplings ${\cal L}^{\rm scal}$, ${\cal L}^{\rm vec}$ and,
in the following, we will not make use of the specific form of the scalar potential (\ref{potential}) and the kinetic matrix (\ref{MJ10}) dictated by superconformal invariance.
The topological term on the other hand is universal with its form determined by the non--Abelian tensor hierarchy of
Section~\ref{sec:hierarchy}.

\subsection{Action for chiral tensor fields}
\label{subsec:ct}

Let us now describe in detail the chiral tensor field Lagrangian entering the actions (\ref{miniL}), (\ref{Lext}). It has the following form
\begin{eqnarray}\label{L6HH=}
{\cal L}^{\mathcal{HH}} = - ( i_v* {\cal H}^I_3+ {i_v\cal H}^{I}_{3}) \wedge {\cal H}_{3I}\wedge v = {1\over 2}
d^6 x \,
 v^\rho\left(*{\cal H}^{I}_{\mu\nu\rho} +     {\cal
H}^{I}_{\mu\nu\rho} \right)(*{\cal H}_I^{\mu\nu\lambda})\,v_\lambda \; , \qquad
\end{eqnarray}
where the one--form
\begin{eqnarray}\label{v1=}
v:= dx^\mu   v_\mu  = \frac{dx^\mu\partial_\mu a(x)}{\sqrt{\partial_\mu a \partial^\mu a}}\; , \qquad v_\mu
v^\mu=1 \; ,
\end{eqnarray}
is the normalized derivative of the auxiliary scalar field $a(x)$, whose presence in the action ensures
 its space--time covariance
 (see \cite{Pasti:1996vs} for the Abelian chiral field case in $D=6$). Consistency of the construction requires
 that the action~\eqref{miniL} is invariant under a local symmetry which allows one to gauge fix $v_\mu$
 to a constant value and moreover that the
variation of the action produces the desired equations of motion. To this end, let us consider a generic variation
of \eqref{miniL} with respect to the scalar and tensor fields.
The variation of the Lagrangian (\ref{L6HH=}) reads
\begin{eqnarray}\label{vL6HH=}
\delta {\cal L}^{\mathcal{HH}} &=&  2  i_v \left( *{\cal H}^{I}_{3} +  {\cal H}^I_3\right) \wedge v \wedge \left(
\delta {\cal H}^I_3 - {1\over 2} \delta v\wedge \left( i_v  *{\cal H}^{I}_{3} +  i_v {\cal H}^I_3\right) \right) -
{\cal H}_{3 \, I}\wedge\delta {\cal H}^I_3   \; , \qquad
\end{eqnarray}
where $\delta {\cal H}^I_3$ was defined in \eqref{Delta2} and
\begin{eqnarray}\label{vv1=}
\delta v = dx^\mu \delta v_\mu \; ,  \qquad \delta v_\mu = \frac{(\eta_{\mu\nu}- v_\mu v_\nu)\partial^\nu \delta
a(x)}{\sqrt{\partial_\lambda a\partial^\lambda a } } \; . \qquad
\end{eqnarray}
 To obtain (\ref{vL6HH=}), the following identities are useful
\begin{eqnarray}\label{v-id}
F_p\equiv i_vF_p\wedge v + *(i_v*F_p\wedge v) \; , \qquad F_6=i_v F_6\wedge v
 \; ,  \qquad i_v * {\cal H}^{I}_{3} \equiv * ({\cal H}^{I}_{3} \wedge  v) \;,
\end{eqnarray}
and $F_p\wedge *G_p= G_p\wedge *F_p$.  Introducing the notation
\begin{eqnarray}\label{cG2I=}
{\cal G}_2^I:= {i_v({ *{\cal H}^{I}_{3} +   {\cal H}^I_3)}\over \sqrt{\partial a
\partial a}}
\;,
\end{eqnarray}
we can write (\ref{vL6HH=}) as
\begin{eqnarray}\label{vL6HH=da}
\delta {\cal L}^{\mathcal{HH}} &=&  2 {\cal G}_2^J\eta_{IJ} \wedge da \wedge  \left( \delta {\cal H}^I_3 - {1\over
2} d(\delta a)\wedge {\cal G}_2^I  \right) - {\cal H}_{3 \, I}\wedge\delta {\cal H}^I_3   \; , \qquad
\end{eqnarray}
Now, using equations  \eqref{Delta2} and the Bianchi identities (\ref{DcF=}), (\ref{DcHI=}), one  gets
\begin{eqnarray}\label{vL6HH=nAb}
\delta {\cal L}^{\mathcal{HH}}  &=&  2 {\cal G}_{2I} \wedge da \wedge  \left( D\Delta {B}^I_2 - {1\over 2}d(\delta
a)\wedge \,{\cal G}_2^I +2 d_{st}^I {\cal F}^s\wedge \delta A^t  + g^{Ir} \Delta C_{3r}\right)+  \nonumber \\ && +
\left( g_I^r {\cal H}_{4r} + d_{Ist} {\cal F}^s\wedge {\cal F}^t \right)\wedge  \Delta {B}^I_2 - g_I^r {\cal
H}_{3}^I  \wedge  \Delta {C}_{3r} - 2d_{Ist} {\cal H}_{3}^I  \wedge {\cal F}^s \wedge \delta A^t\; . \qquad
\end{eqnarray}
 Terms similar to those in the second line of \eqref{vL6HH=nAb} enter the variation of the topological term
${\cal L}^{\rm top}$ \eqref{dLtop=}
\begin{eqnarray}\label{vLtop=}
\delta  {\cal L}^{\rm top}  &=& \; \left( -d_{Ist} {\cal F}^s\wedge {\cal F}^t + g_I^r {\cal H}_{4r} \right)\wedge
\Delta {B}^I_2 + g_I^r {\cal H}_{3}^I \wedge \Delta {C}_{3r} - 2d_{Ist} {\cal H}_{3}^I \wedge {\cal F}^s \wedge
\delta A^t\; . \qquad
\end{eqnarray}
 Thus
\begin{eqnarray}\label{vLHH+cLt=}
\delta\left( {\cal L}^{\mathcal{HH}} +  {\cal L}^{\rm top}\right)   &=&  2 {\cal G}_2^J\eta_{JI} \wedge da \wedge
\left( D\Delta {B}^I_2 - {1\over 2}(\delta a)\,{\cal G}_2^I +2 d_{st}^I {\cal F}^s\wedge \delta A^t + g^{Ir}
\Delta C_{3r} \right)+ \nonumber \\ &&+ 2 g_I^r {\cal H}_{4r} \wedge  \Delta {B}^I_2 -4d_{Ist} {\cal H}_{3}^I
\wedge {\cal F}^s \wedge \delta A^t\; . \qquad
\end{eqnarray}
Combining this with the variation of the matter Lagrangian ${\cal L}^{\rm scal}+{\cal L}^{\rm vec}$, we finally
obtain the variation of the full Lagrangian \eqref{miniL}
\begin{eqnarray}\label{vL+L+L=}
\delta {\cal L} & =& 2 {\cal G}_2^J\eta_{JI} \wedge da \wedge \left( D\Delta {B}^I_2 - {1\over 2}(\delta a)\,{\cal
G}_2^I +2 d_{st}^I {\cal F}^s\wedge \delta A^t + g^{Ir} \Delta C_{3r} \right)+ \nonumber \\ &&+ 2\,g_I^r \left(
{\cal M}_{rs} * {\cal F}^s+ {\cal H}_{4r} \right) \wedge  \Delta {B}^I_2 +\left( {\cal J}_t + 2 D( {\cal M}_{ts}
*{\cal F}^s) - 4\,d_{Ist} {\cal H}_{3}^I  \wedge {\cal F}^s \right) \wedge \delta A^t \nonumber \\[1.5ex] &&+
\delta_Y {\cal L} + \delta_\phi {\cal L}\;,
\end{eqnarray}
with the matter current ${\cal J}_r$ defined by the variation of the matter Lagrangian as
\bea \delta{\cal L}^{\rm scal} &=& {\cal J}_r \wedge \delta A^r ~\equiv~ k_r^\alpha {\cal J}_{\alpha} \wedge \delta A^r
\;.
\eea

The form of eq.~\eqref{vL+L+L=} suggests that the action \eqref{miniL} is invariant under the following local
transformations of $B^I_2$ and $C_{3r}$
\begin{eqnarray}\label{PST1=nonAb}
\Delta_{\varphi_1}  B_{2}^I =   \varphi^I_{1} \wedge da\;, \qquad \Delta_{\varphi_2} C_{3r} =   \varphi_{2r}
\wedge da  \; , \;  \;  \qquad
\end{eqnarray}
where the  two--form parameter $ \varphi_{2r}(x)$ is arbitrary and the one--form parameter
 $\varphi^I_{1}(x)$ satisfies the condition $g^s_I\varphi^I_1=0$.
Note that in general these symmetries are not included in the tensor gauge  symmetries of (\ref{gaugesym}) whose
parameters $\Lambda_{2\,r}$ and $\Lambda_{3\,\alpha}$
appear only under the projection with the tensors $g^{Ir}$ and
$k_r^\alpha$, respectively.

Another local symmetry of the action is the one which exposes the auxiliary nature of the scalar field $a(x)$
\begin{eqnarray}\label{PST2=nonAb}
 \delta a=\varphi(x)\,,\qquad \Delta_{\varphi} {B}^I_2 = \delta a\,{\cal G}_2^I  \; ,
 \qquad   \Delta_{\varphi}
C_{3r} = \, \delta a\, {\cal G}_{3r} \; , \qquad \end{eqnarray} where $\varphi(x)$ is an arbitrary scalar
parameter and
\begin{eqnarray}\label{cG3r=}
{\cal G}_{3r}:= {i_v({\cal H}_{4r}+{\cal M}_{rs}*{\cal F}^s)\over \sqrt{\partial a \partial a}}\, .
\end{eqnarray}
One can use this symmetry to gauge fix $v_\mu$ to be e.g. the constant unit time--like vector \be\label{gf}
v_\mu=\delta^0_\mu\,. \ee If in \eqref{L6HH=} we substitute $v_\mu$ with its gauge fixed value \eqref{gf}, the
manifest space--time invariance of the action will be broken and it reduces to the non--Abelian generalization of
the Henneaux--Teitelboim action \cite{Henneaux:1987hz,Henneaux:1988gg} for a single chiral 2--form in $D=6$.
However, the gauge--fixed action is still invariant under modified Lorentz transformations, which preserve the
gauge \eqref{gf}. They are the combination of Lorentz rotations with the parameters $l_\mu{}^\nu$ and local
transformations \eqref{PST2=nonAb} and \eqref{vv1=} such that \be\label{modi} \Delta_{L}
v_\mu=\delta_lv_\mu+\delta_\varphi v_\mu =\Delta_{L}
(\delta^0_\mu)=0=l_\mu{}^0+\partial_\mu\varphi-\delta^0_\mu\partial_0\varphi\, \ee from which it follows that
\be\label{vp} \varphi(x)=-x^\mu\l_\mu{}^0 \ee and the modified Lorentz transformations under which the gauge fixed
action is invariant are \be\label{modL} \Delta_{L} {B}^I_2 = \delta_l {B}^I_2-x^\mu\l_\mu{}^0\,{\cal G}_2^I  \; ,
 \qquad   \Delta_{L}
C_{3r} =\delta_l C_{3r}-\,x^\mu\l_\mu{}^0 {\cal G}_{3r}. \ee In \eqref{modL} it is implied that in the quantities
${\cal G}_2^I$ and ${\cal G}_{3r}$, defined in \eqref{cG2I=} and \eqref{cG3r=}, $v_\mu$ takes its gauge fixed
value \eqref{gf}.

\subsubsection{Derivation of the field equations}
\label{subsec:eomtensor}

Let us now discuss the derivation of the field equations from the variation \eqref{vL+L+L=} of the action. It
demonstrates in an instructive manner how the tensor hierarchy intertwines equations of motion of different tensor
fields. For the analysis of the equations of motion it is useful to introduce a constant
projector $\mathbb{P}^{I}{\!}_{J}$ of minimal rank satisfying
\bea g^r_I\,\mathbb{P}^{I}{\!}_{J} = g^r_J \;,\qquad
\mathbb{P}^{I}{\!}_{J}\mathbb{P}^{J}{\!}_{K}= \mathbb{P}^{I}{\!}_{K} \; \label{projector}
\eea
and
the complementary (orthogonal) projector
\be
\label{tildeprojector} \bar{\mathbb{P}} = {\rm I} - \mathbb{P}\,,\qquad \bar{\mathbb{P}}\bar{\mathbb{P}}=\bar{\mathbb{P}}\,.
\ee
which obeys $g^r_I\,\bar{\mathbb{P}}^{I}{\!}_{J}=0$. We stress that the introduction of this projector is an auxiliary structure
in order to derive the different parts of the equations of motion,
whereas eventually the combined set of equations of motion does not carry any reference to this projector.

We start with the equation of motion produced by the variation of $C_{3r}$
\be\label{h31} {\cal G}_2^I g^r_I \wedge da
:=g^r_I\,i_v(\mathcal H_3^I+*\mathcal H_3^I)\wedge v=0\,. \ee
Due to the properties of the projector \eqref{projector}  and its complementary \eqref{tildeprojector}, we see that this equation is equivalent to
\be\label{h310}
\mathbb P^{I}{}_{J}\,i_v(\mathcal H_3^J+*\mathcal H_3^J)\wedge v=0,
\ee
since \eqref{h31} is satisfied if and only if \eqref{h310} holds.
In view of the identities
\eqref{v-id},  eq.~\eqref{h310} is amount to the anti--self--duality of the part of $H_3^I$ projected with $\mathbb P^I{}_J$
 \be\label{h32p}
 \mathbb{P}^I{\!}_{J}
 \left(\mathcal H_3^J+*\mathcal H_3^J\right)=0\;.
\ee
Moreover, we can use the second symmetry in \eqref{PST1=nonAb} to put
\be\label{h32p2}
i_v(\mathcal H_{3I}+*\mathcal H_{3I})\wedge v\,{\mathbb P}^I{}_{J}=0\qquad \Longrightarrow\qquad
(\mathcal H_{3I}+*\mathcal H_{3I})\,{\mathbb P}^I{}_{J}=0\,.
\ee
Indeed, under  the second symmetry in \eqref{PST1=nonAb} $\delta {\cal G}_{2I}= g_I^r\varphi_{2r}\,\, \Rightarrow \,\, \delta {\cal G}_{2J} {\mathbb P}^J{}_{I}=g_I^r\varphi_{2r}$, which can be used to fix $ {\cal G}_{2J} {\mathbb P}^J{}_{I}=0$.
Now, the variation of $\Delta B_2^I$ gives
\be \label{h33} D\left(i_v(\mathcal H_3^I+*\mathcal
H_3^I)\wedge v\right)-g^{Ir}({\cal H}_{4r}+ {\cal M}_{rs} *{\cal F}^s) =0\;. \ee
Projecting the above equation with $\mathbb P^I{}_{J}$ and $\bar{\mathbb P}^I{}_{J}$, in view of \eqref{h32p2} we get
\bea
 \label{h331}
g^{r}_J({\cal H}_{4r}+ {\cal M}_{rs} *{\cal F}^s) &=&0\;,\\
\label{h3311}
\bar{\mathbb{P}}^I{}_J{} \,D\left(i_v(\mathcal H_{3I}+*\mathcal
H_{3I})\wedge v\right)&=&0\,.
\eea
The equation \eqref{h331} is a projected
version of the duality relation between $\mathcal H_{4r}$ and $\mathcal F^r$. As for equation (\ref{h3311}), it
reduces to
\bea \bar{\mathbb{P}}^I{}_J \,d\left(i_v(\mathcal H^J_{3}+*\mathcal H^J_{3})\wedge v\right)  &=& 0,
\label{dH} \eea
since, by virtue of (\ref{h32p}), the non-trivial connection part of the covariant derivative $D$ in (\ref{h3311}) is
\bea
\bar{\mathbb{P}}^K{}_I  \, X_{r\,LK} \left(i_v(\mathcal H_{3}^L+*\mathcal H_{3}^L)\wedge v\right)  &=&
\bar{\mathbb{P}}^K{}_I \bar{\mathbb{P}}^L{}_J \, X_{r\,LK} \left(i_v(\mathcal H_{3}^J+*\mathcal H_{3}^J)\wedge
v\right)\;,
\eea
and thus vanishes since
\bea \bar{\mathbb{P}}^K{}_I
\bar{\mathbb{P}}^L{}_J \, X_{r\,LK}&=&0 \;,
\eea
according to the definition of $X_{r\,L}{}^K$ in
\eqref{DcHI=d+} and of the projectors in (\ref{projector}) and (\ref{tildeprojector}). Then, equation~(\ref{dH})
can be solved in the same way as in the case of the Abelian chiral tensor fields \cite{Pasti:1996vs} with the
general solution (at least locally or in the topologically trivial cases) being
\be\label{h37} i_v(\mathcal
H_3^I+*\mathcal H_3^I)\wedge v=d (\phi^I_1\wedge da)\,, \ee where the one--form $\phi^I_1(x)$ is such that
$\phi^I_1 g_I^r=0$,  or equivalently $\mathbb{P}^K{}_I \phi^I_1 =0$. One can now use the local symmetry \eqref{PST1=nonAb} with the parameter $\varphi^I_1(x)$ (also obeying $\varphi^I_1 g_I^r=0$) to
annihilate the right hand side of eq. \eqref{h37} and, in view of \eqref{h32p}, arrive  at the anti--self--duality
condition for all $\mathcal H^I_3$
\be \label{h38} \mathcal H_3^I+*\mathcal H_3^I=0\;.
\ee
 When \eqref{h38} is satisfied, the variation \eqref{vL+L+L=} with respect to $\delta a$ vanishes identically, thus confirming that the scalar $a(x)$ is entirely auxiliary, while the variation of $A^r$ provides us with the vector field equations of
motion. The complete set of the field equations obtained from varying the action \eqref{miniL} with respect to the
$p$-forms is  \bea\label{completeset} \mathcal H_3^I+*\mathcal H_3^I&=&0\;,\nonumber\\ g^{Ir}({\cal H}_{4r}+{\cal
M}_{rs}*{\cal F}^s) &=&0\;,\nonumber\\
 2 D({\cal M}_{st}*{\cal F}^s)+ {\cal J}_t - 4d_{Ist} {\cal H}_{3}^I  \wedge
 {\cal F}^s
 &=&0\;.
 \eea
 Notice that the field $a(x)$ does not enter these equations. This once again manifests the fact that $a(x)$ is
completely auxiliary and is only required for ensuring the space--time covariance of the action.

The (bosonic limit of the) (1,0)
models of~\cite{Samtleben:2011fj} are recovered with the particular choice of $\mathcal M_{rs}$ and $J_t$ as in eq.(\ref{MJ10})
and the scalar potential as in (\ref{potential}) dictated by supersymmetry. In this case the variation of the action \eqref{miniL} with respect to the scalar fields yields the equations of
motion  \bea
 D*D\phi_I-2d_{Irs}{\cal F}^r \wedge  *{\cal
F}^s-d^6x\,(2d_{Irs}Y^{ij\,r}Y_{ij}^s+3g^r_{(J}g^s_{K}d_{I)rs}\phi^J\phi^K)=0\;,\nonumber\\ \phi^I
d_{Irs}Y^{ij\,r}=0\;. \eea
Comparing (\ref{completeset}) to the full system of first-order duality equations (\ref{general-selfdual}), we see that the
action (\ref{miniL}) gives rise to the full self-duality equation (\ref{HH}) but only to a projection of the
duality equation between the vector and the three--form gauge potentials. This will be rectified in the next section.

Finally, before concluding this section let us note that, using the Bianchi identities (\ref{DcH4=}) one can
rewrite the general variation \eqref{vL+L+L=} as follows
 \begin{eqnarray}\label{vL+L+L=vADK4+}
\delta {\cal L} &=&  2 {\cal G}_2^J\eta_{JI} \wedge da \wedge  \left( D\Delta {B}^I_2 - {1\over 2}d(\delta
a)\wedge {\cal G}_2^I +2 d_{st}^I {\cal F}^s\wedge \delta A^t  + g^{Ir} \Delta C_{3r} \right) \nonumber
\\ &&+ 2
g_I^r {\cal K}_{4r} \wedge  \Delta {B}^I_2 +  2 D{\cal K}_{4r}\wedge \delta A^r  + k_r^\alpha\left( *{\cal J}_\alpha-2\,
{\cal H}_{5\alpha}\right) \wedge \delta A^r \nonumber \nonumber \\[1.5ex] &&+ \delta_Y {\cal L} + \delta_\phi
{\cal L}\; .
\end{eqnarray}
where
\begin{eqnarray}\label{cK4=}
{\cal K}_{4\,r} := {\cal H}_{4r}+ {\cal M}_{rs}*{\cal F}^s  \;  . \qquad
\end{eqnarray}
The last term in the second line of \eqref{vL+L+L=vADK4+} infers that the four-form gauge potential $k_r^\alpha C_{4\alpha}$ can be dual to the
scalars $\phi^I$ (see eq. \eqref{H5}). This duality condition, however, does not follow from the above action
(note that the action \eqref{miniL} does not even contain the four-form field $C_{4\alpha}$).
In the next section, we will construct an action for the extended tensor hierarchy system, that explictly
includes the four-form $C_{4\alpha}$ and also treats the vector and three--form fields $A^r_1$ and $C_{3r}$ in a
duality-symmetric fashion. Equation~(\ref{H5})
will then appear as a full--fledged equation of motion.

\subsection{Action with manifest vector-tensor duality symmetry}
\label{subsec:vt}

In this section we will extend the action (\ref{miniL}) to the form (\ref{Lext})
in such a way that it treats the vector and three--form tensor
fields on an equal footing and yields their first-order duality relation (\ref{cK4=})
among the proper equations of motion.
The corresponding action includes all the $p$-form fields $(Y^{ij\,r},\phi^I,A^r_1,B_2^I,C_{3\,r},C_{4\alpha})$
and is obtained by adding to the action \eqref{miniL} the following term
\begin{eqnarray}\label{LH4F=}
{\cal L}^{{\cal H}_4/{\cal F}}= - {1\over 4}\, \tilde{\mathcal M}^{rs}\, (i_v *{\cal K}_{4r}) \wedge * (i_v *{\cal
K}_{4s}) \; . \qquad
\end{eqnarray}
where ${\cal K}_{4r}$ has been defined in \eqref{cK4=}, and the matrix
 $\tilde{\mathcal M}^{rs}$ is such that
\be\label{tildephi} \tilde{\mathcal M}^{rs}{\mathcal M}_{st}=P^r{}_s,\qquad
\tilde{\mathcal M}^{rs}{\mathcal M}_{st}\tilde{\mathcal M}^{tq}=\tilde{\mathcal M}^{rq},\qquad
{\mathcal M}_{st}\tilde{\mathcal M}^{tq}{\mathcal M}_{qr}={\mathcal M}_{sr}\,, \ee
where $P^r{}_s$ is the
projector of the same rank as ${\mathcal M}_{st}$, i.e.
$$
{\mathcal M} P={\mathcal M}\,, \qquad P\tilde{\mathcal M}=\tilde{\mathcal M}\,.
$$
If ${\mathcal M}_{rs}$ is invertible, which is the case we shall mostly deal with,
$\tilde{\mathcal M}^{rs}$ is inverse of ${\mathcal M}_{rs}$, \emph{i.e. }
\be\label{inverse}
P^r{}_s=\tilde{\mathcal M}^{rt}{\mathcal M}_{ts}=\delta^r_s\,.
\ee

The full duality--symmetric Lagrangian is given by (\ref{Lext}).
The terms ${\cal L}^{\rm vec}$ from \eqref{LcF*cF=} and ${\cal L}^{{\cal H}_4/{\cal F}}$ from \eqref{LH4F=}
together form the duality--symmetric Lagrangian for the fields $A_1^r$
and $C_{3r}$. Indeed, their sum can be rewritten in the
following manifestly duality-symmetric form \bea \label{ds}  {\cal L}^{\rm vec}+{\cal L}^{{\cal
H}_4/{\cal F}}&=&\frac{1}{2}{\mathcal M}_{rs} {\cal F}^r\wedge *{\cal F}^s-\frac{1}{2}\tilde{\mathcal M}^{rs} {\cal H}_{4r}\wedge
*{\cal H}_{4s} -(\tilde{{\mathcal M}}{{\mathcal M}})^r{}_s\, {\cal H}_{4r}\wedge \mathcal F^s
\nonumber\\ &&{}+ {1\over 4}\, \tilde{{\mathcal M}}^{rs}\, (i_v *{\cal K}_{4r}) \wedge * (i_v *{\cal
K}_{4s})+ \frac{1}{2}\tilde{{\mathcal M}}^{rs}\, (i_v{\cal K}_{4r}) \wedge * (i_v {\cal
K}_{4s})\;. \eea

We should now check that the addition of the Lagrangian \eqref{LH4F=} to the action \eqref{miniL} does not spoil
the local symmetries \eqref{PST1=nonAb} and \eqref{PST2=nonAb}. Using the relation $*(i_v *{\cal K}_{4r}\wedge v)=
{\cal K}_{4r}-  i_v{\cal K}_{4r}\wedge v$ we find that the generic variation of \eqref{LH4F=} is
\begin{eqnarray}\label{vLH4F=}
\delta {\cal L}^{{\cal H}_4/{\cal F}}&=& -\frac 12 \tilde{{\mathcal M}}^{rs}\,  {\cal G}_{3r} \wedge {\cal G}_{1s}\wedge
da \wedge d\delta a  +\frac 12 \tilde{{\mathcal M}}^{rs}\,  {\cal G}_{1s}\wedge da \wedge \delta {\cal H}_{4r}
   \nonumber \\ && -\frac 12 \tilde{{\mathcal M}}^{rt}\, {{\mathcal M}}_{ts}\,  {\cal G}_{3r} \wedge
  da \wedge \delta {\cal F}^s
  +\frac 12 \tilde{{\mathcal M}}^{rt}\, {{\mathcal M}}_{ts}\,  {\cal K}_{4r} \wedge \delta {\cal F}^s
  \nonumber \\ && + \frac 14\, \delta\tilde{{\mathcal M}}^{rs}\,  i_v (*{\cal K}_{4r})
  \wedge {\cal K}_{4s} \wedge v_1 +\frac 12\, \tilde{{\mathcal M}}^{rt} \delta {{\mathcal M}}_{ts}\,
  {\cal F}^s \wedge {\cal K}_{4r}  \; ,   \qquad
\end{eqnarray}
where we introduced the definitions
\begin{eqnarray}\label{cG1=}
\label{cG1r=} && {\cal G}_{1r}:= {i_v(*{\cal K}_{4r})\over \sqrt{\partial a \partial a}} \; , \qquad {\cal
G}_{3r}:= {i_v{\cal K}_{4r}\over \sqrt{\partial a \partial a}} \;, \qquad
\end{eqnarray}
in accordance with (\ref{cG3r=}).
Adding this variation to \eqref{vL+L+L=vADK4+} and making use of
the explicit form of $\delta \mathcal F^r$ and
$\delta \mathcal H_{4r}$ given in \eqref{Delta2} we get
 \bea\label{deltaLex}
  \delta {\cal L}_{\rm ext} &=&  2 {\cal
G}_2^J\eta_{JI} \wedge da \wedge  \Big( D\Delta {B}^I_2 - {1\over 2}d(\delta a)\wedge {\cal G}_2^I +2 d_{st}^I
{\cal F}^s\wedge \delta A^t  + g^{Ir} \Delta C_{3r} \Big) \nonumber
\\ &&{}+
 2\tilde{{\mathcal M}}^{rs}\,  {\cal G}_{3r} \wedge {\cal G}_{1s}\wedge da \wedge d\delta a -
 2 \tilde{{\mathcal M}}^{rs}\,  {\cal G}_{1s}\wedge da \wedge D\Delta {C}_{3r}
\nonumber
\\ &&{}+ 2 g_I^s(\delta-\tilde{\mathcal M}{\mathcal M})^r{}_s {\cal K}_{4r} \wedge  \Delta {B}^I_2 +  2
(\delta-\tilde{\mathcal M}{\mathcal M})^r{}_s{\cal K}_{4r}\wedge D\delta A^s
+ k_r^\alpha\left(* {\cal J}_\alpha- 2 \, {\cal H}_{5\alpha}\right)\wedge \delta A^r
\nonumber
\\ &&{} +  4 \tilde{{\mathcal M}}^{rt}\,  {\cal
G}_{1t}\wedge da \wedge {\cal F}^sd_{Irs} \wedge \Delta B_2^I
  + 2 \tilde{{\mathcal M}}^{rt}\, {{\mathcal M}}_{ts}\,  {\cal G}_{3r} \wedge da \wedge   g^s_I\Delta
  B_2^I
\nonumber
\\
 &&{} +  4 \tilde{{\mathcal M}}^{rt}\,  {\cal G}_{1t}\wedge da \wedge {\cal H}_{3}^I \wedge
 \delta A^s d_{Irs}
  + 2 \tilde{{\mathcal M}}^{rt}\, {{\mathcal M}}_{ts}\,  {\cal G}_{3r} \wedge da \wedge  D\delta A^s
 \nonumber
 \\
&&{}  -  2 \tilde{{\mathcal M}}^{rs}\,  {\cal G}_{1s}\wedge da \wedge k_r^\alpha \Delta C_{4\alpha}
+\delta_Y \,{\cal L} +\delta_{\phi} \,{\cal L}
 \;.
 \eea
 One can check
that this variation vanishes for the local symmetry transformations (\ref{PST2=nonAb}) provided that $A_1^r$ and
$C_{4\alpha}$ transform as follows
\bea\label{PST2==nonAb}
\delta A^{r}&=& \, \delta a \,  \tilde{{\mathcal M}}^{rs}{\cal G}_{1s} \; , \;  \qquad
\\[1ex]
 \Delta{C}_{4\alpha}k^\alpha_r&=& \frac{\delta a}{\sqrt{(\partial a)^2}}\, \left(k_r^\alpha i_v
{\cal H}_{5\alpha}-2g^s_{[I}d_{J]sr}\,{\mathcal M}^I i_v *D{\mathcal M}^J + *({\cal G}_{1t}\wedge da)\,
i_vD(\tilde{{\mathcal M}}{\mathcal M})^t{}_r \right)\nonumber\\
&&{}
-(\delta-\tilde{\mathcal M}{\mathcal M})^s{}_{r}X_{4s}\;,
\label{kvC41} \eea where the four--form $X_{4s}$ is
such that
$$
X_{4s}(\delta-\tilde{\mathcal M}{\mathcal M})^s{}_{r}g^{Ir}=\frac{\delta a}{\sqrt{(\partial a)^2}}\,g^{Ir} *({\cal G}_{1t}\wedge
da)\, i_vD(\tilde{{\mathcal M}}{\mathcal M})^t{}_r \;.
$$
This relation has solutions when $(\tilde{\mathcal M}{\mathcal M})^{t}{}_{r}g^{Ir}=0$. It is trivially satisfied in the case of non--degenerate ${\mathcal M}_{rs}$, i.e.\ when $(\tilde{\mathcal M}{\mathcal M})^{t}{}_{r}=\delta^{t}{}_{r}$. It is this case that we shall consider in detail in the following. If on the other hand ${\mathcal M}_{rs}$ is degenerate, some vector gauge fields do not have the kinetic terms in the Lagrangian and are therefore non--dynamical.
In~\cite{Samtleben:2012fb} it has been shown that for the
$(1,0)$ superconformal models, invertibility of ${\cal M}_{rs}$ from (\ref{MJ10}) can always be achieved by
including Abelian factors in the gauge group.

\subsubsection{Derivation of the field equations}

Let us now discuss the derivation of the field equations from the variation (\ref{deltaLex})
of the extended Lagrangian (\ref{Lext}), assuming that the kinetic matrix ${\cal M}_{rs}$ of the vector fields is invertible \eqref{inverse}.
In this case, the variation (\ref{deltaLex}) reduces to
 \bea\label{deltaLexI}
  \delta {\cal L}_{\rm ext} &=&  2 {\cal
G}_2^J\eta_{JI} \wedge da \wedge  \Big( D\Delta {B}^I_2 - {1\over 2}d(\delta a)\wedge {\cal G}_2^I +2 d_{st}^I
{\cal F}^s\wedge \delta A^t  + g^{Ir} \Delta C_{3r} \Big) \nonumber
\\ &&{}+
 2{\tilde{\cal M}}^{rs}\,  {\cal G}_{3r} \wedge {\cal G}_{1s}\wedge da \wedge d\delta a -
 2 {\tilde{\cal M}}^{rs}\,  {\cal G}_{1s}\wedge da \wedge D\Delta {C}_{3r}
  + 2 \,  {\cal G}_{3r} \wedge da \wedge   g^r_I\Delta B_2^I
\nonumber
\\ &&{} +  4 {\tilde{\cal M}}^{rt}\,  {\cal
G}_{1t}\wedge da \wedge {\cal F}^sd_{Irs} \wedge \Delta B_2^I
+  4 {\tilde{\cal M}}^{rt}\,  {\cal G}_{1t}\wedge da \wedge {\cal H}_{3}^I \wedge
 \delta A^s d_{Irs}
 \nonumber
\\
 &&{}
  + 2 \,  {\cal G}_{3r} \wedge da \wedge  D\delta A^r
+ k_r^\alpha\left(* {\cal J}_\alpha- 2 \, {\cal H}_{5\alpha}\right)\wedge \delta A^r
  -  2 {\tilde{\cal M}}^{rs}\,  {\cal G}_{1s}\wedge da \wedge k_r^\alpha \Delta C_{4\alpha}
 \nonumber
 \\
&&{} +\delta_Y \,{\cal L} _{\rm ext}+\delta_\phi \,{\cal L}_{\rm ext}
 \;.
 \eea
It follows that this variation vanishes under an extension of the local symmetry transformations (\ref{PST1=nonAb}) and is invariant under
\be\label{PST1ext}
\delta A^r = \varphi^r \,  da\;,\quad
\Delta_{\varphi_1}  B_{2}^I =   \varphi^I_{1} \wedge da\;, \quad
\Delta_{\varphi_2} C_{3r} =   \varphi_{2r}
\wedge da \;,\quad
\Delta C_{4\alpha}=\varphi_{3\alpha}\wedge da\,,
\ee
where the parameters $\varphi^I_{1}$, $\varphi_{2r}$, and $\varphi_{3\alpha}$ are arbitrary
and $\varphi^r$ satisfies $k_r^\alpha\,\varphi^r=0$\,.
Similar to (\ref{projector}) it turns out to be useful to introduce two projectors ${\mathbb P}_r{}^s$ and ${\mathcal P}^r{}_s$
of minimal rank satisfying
\bea
k^\alpha_r\,{\mathcal P}^r{}_s &=& k_s^\alpha\;,
\qquad
g_I^r\,{\mathbb P}_r{}^s ~=~ g_I^s
\;,
\eea
respectively, together with their respective complementary projectors defined according to (\ref{tildeprojector}).
The orthogonality $g_I^r\,k_r^\alpha =0$ implies that
\bea
{\mathcal P}^r{}_t\,{\mathbb P}_s{}^t &=& 0\;,
\eea
whereas the opposite contraction of the two projectors is not necessarily vanishing.
The equations of motion which follow from the $\Delta C_{4\alpha}$ variation of (\ref{deltaLexI}) are
 \bea\label{C4alEqI}
 && k_r^\alpha\, {\tilde{\cal M}}^{rs}\,{\cal G}_{1s}\wedge da=0\; . \qquad
  \eea
By construction ${\cal G}_{1s}\propto i_v {\cal K}_4$ does not contain any contribution proportional to $da$,
 which means that  (\ref{C4alEqI}) implies
  \bea\label{kcG1=0I}
 && {\mathcal P}^r{}_t \, {\tilde{\cal M}}^{ts}\,{\cal G}_{1s}=0\; . \qquad
 \eea
Let us turn to the equations appearing as the coefficient for $\Delta C_{3r}$ in the variation (\ref{deltaLexI}) :
\bea\label{C3rEqI}
 &&  D ({\tilde{\cal M}}^{rs}{\cal G}_{1s}\wedge da)= g^{Ir} {\cal G}_{2I}\wedge da \; . \qquad
 \eea
Upon projection with $\bar{\mathbb P}$, we find
\bea
0 &=& \bar{\mathbb P}_s{}^r\,D ({\tilde{\cal M}}^{st}{\cal G}_{1t}\wedge da)
~=~ \bar{\mathbb P}_s{}^r\,\bar{{\mathcal P}}^s{}_v\,d \left({\tilde{\cal M}}^{vt}{\cal G}_{1t}\wedge da\right)
\;,
\eea
where the second equality uses (\ref{C4alEqI}) and the fact that the connection part vanishes due to
\bea
\bar{\mathbb P}_s{}^r\,\bar{{\mathcal P}}^t{}_v \,X_{u\, t}{}^s &=&
 \bar{\mathbb P}_s{}^r\,\bar{{\mathcal P}}^t{}_v \,(k_t^\alpha c_{\alpha u}^s+2d^I_{tu}g_I^s)~=~
0\;.
\label{zeroproj}
\eea
Similar to the Abelian case, we thus conclude that locally
\bea
\left(\bar{\mathbb P}_s{}^r\, {\tilde{\cal M}}^{st}{\cal G}_{1t} \right) \wedge da &=& d\left(\bar{\mathbb P}_s{}^r\, \phi^s \; da \right)
\;,
\eea
with $\phi^s$ satisfying $k_s^\alpha \phi^s=0$\,.
We can thus use the local symmetry (\ref{PST1ext}) with the parameter $\varphi^r$ (also obeying $k_s^\alpha \varphi^s=0$)
to obtain $\bar{\mathbb P}_s{}^r\, {\tilde{\cal M}}^{st}{\cal G}_{1t}=0$\,.
Finally, the local symmetry (\ref{PST1ext}) with properly chosen parameter $\varphi^I_1$
can be used to extend this equation to the full duality equation
\bea\label{g1}
{\tilde{\cal M}}^{rs}\,{\cal G}_{1s}=0
\;.
\eea
We note, that this fixes the local symmetry with parameter $\varphi^I_1$ up to parameters satisfying
$g_I^r \varphi^I_1=0$ which do not contribute to the variation of ${\tilde{\cal M}}^{rs}\,{\cal G}_{1s}$.
We are thus left with the local symmetries of (\ref{PST1=nonAb}) above, while eq. (\ref{C3rEqI}) reduces to \eqref{h31}. Thus we can proceed as in Section~\ref{subsec:eomtensor}
for the minimal case and obtain
\bea\label{gHI+g*HI=0I}
 && g^{Ir} ({\cal H}_{3I}+ *{\cal H}_{3I}) =0  \; . \qquad
 \eea
Let us turn to the equations produced by the variation $\Delta B_{2}^I$ in (\ref{deltaLexI}). In view of \eqref{g1} we get
\bea\label{vB2IEqI}
 D ({\cal
G}_{2I} \wedge da)=\, g^r_I \,{\cal G}_{3r} \wedge da \,.
 \eea
Equations (\ref{gHI+g*HI=0I}) and (\ref{vB2IEqI}) are precisely analogous to
equations (\ref{h32p}), (\ref{h33}) which have been our starting point
in the discussion of field equations in the minimal case in Section~\ref{subsec:eomtensor}.
Proceeding as above, we may thus further gauge fix the remaining local symmetries of (\ref{PST1=nonAb})
and arrive at the field equations
\bea
{\cal H}_{3I}+ *{\cal H}_{3I} &=& 0\;,\qquad
g^r_I \,{\cal G}_{3r} ~=~ 0\;.
\label{eomintI}
\eea

Finally, let us turn to the equations appearing as the coefficient for the vector fields $\delta A^r$ in (\ref{deltaLexI}).
Upon using all field equations that we have already derived, these equations reduce to
\bea
  2 \, D\left( {\cal G}_{3r} \wedge da \right)
&=& k_r^\alpha\left(* {\cal J}_\alpha- 2 \, {\cal H}_{5\alpha}\right)
\;,
\label{DG3}
\eea
and can be solved with the same strategy: projection with $\bar{{\mathcal{P}}}$ yields
\bea
0&=& \bar{{\mathcal{P}}}^r{}_s\,D \left( {\cal G}_{3r} \wedge da \right)  ~=~
\bar{{\mathcal{P}}}^r{}_s\,\bar{\mathbb{P}}_r{}^t\, d \left( {\cal G}_{3t} \wedge da \right)
\;,
\eea
where again we have used (\ref{zeroproj}) together with (\ref{eomintI})
to show that the connection part of the covariant derivative vanishes.
As in the Abelian case we conclude that locally
\bea
\left( \bar{{\mathcal{P}}}^r{}_s\,   {\cal G}_{3r}  \right)\wedge da  &=&
d\left(\bar{{\mathcal{P}}}^r{}_s\, \phi_{3r} \wedge da \right)
\;.
\eea
As above, proper combinations of the remaining local symmetries from (\ref{PST1ext})
allow to obtain ${\cal G}_{3r}=0$\,. Together with (\ref{g1}) we thus obtain ${\mathcal{K}}_{4r}=0$\,.
The r.h.s.\ of (\ref{DG3}) eventually gives the last equation
of (\ref{general-selfdual}).

Summarizing, we have shown that the extended Lagrangian (\ref{Lext}) gives rise to the set of non-Abelian
duality equations (\ref{general-selfdual}).
Again, the field $a(x)$ does not enter these equations, showing that $a(x)$ is
completely auxiliary and is only required for ensuring the space--time covariance of the action.
Via the Bianchi identities (\ref{DcF=}) these equations give rise to the second-order field equations
for the vector fields in (\ref{completeset}).
The last equation in (\ref{general-selfdual}) is a projection of the duality relation (\ref{H5}) between the scalar fields and the four--form gauge fields. In addition, the variation of the Lagrangian (\ref{Lext})
with respect to  the scalar fields gives rise to their standard
second-order field equations.

\section{Example}
\label{sec:example}

Let us now consider an example of a minimal Lagrangian model given in \cite{Samtleben:2012mi}. In this model the
vector fields split into two sets \be\label{A1} A^r=(A^a,\mathcal A^{\hat I}) \ee and the constant tensors
$f_{rs}{}^t$ and $d^I_{rs}$ reduce as follows
\be\label{fd} f_{rs}{}^t\,\rightarrow\,(f_{ab}{}^c,-\frac 12(T_a)_{\hat
I}{}^{\hat J})\,,\qquad d^I_{rs}\,\rightarrow \frac 12(T_a)_{\hat I}{}^{\hat J}\,,
\ee
\emph{i.e., e.g.} $f_{a\hat I}{}^{\hat J}=-f_{\hat I a}{}^{\hat J}=-\frac 12(T_a)_{\hat
I}{}^{\hat J}$,
where the indices $a,b,c$ label the
adjoint representation of a gauge group $G$ whose algebra is defined by the structure constants $f_{ab}{}^c$, and
the indices $\hat I, \hat J$ label representations $\mathcal R$ (upper indices) and $\mathcal R'$ (lower indices)
of $G$ generated by $(T_a)_{\hat I}{}^{\hat J}$.

The scalars $\phi^I$ and the two--form fields $B^I$ split into two sets taking values in $\mathcal R'$ and
$\mathcal R$ \be\label{phiB} \phi^I=(\hat\phi_{\hat I},\phi^{\hat J}),\qquad B^I_2=(\hat B_{2\hat I},B^{\hat
J}_2)\,. \ee It is important to note that the fields with lower and upper indices $\hat I$ are \emph{different}
fields, and that the metric $\eta_{IJ}$ is anti--diagonal \be\label{eta} \eta^{IJ}= \left(
\begin{tabular}{c c}
0 & $\delta^{\hat J}_{\hat I}$ \\
 $\delta^{\hat I}_{\hat J}$ &  0
\end{tabular}
\right)\,. \ee
To be more explicit,  in the case under consideration
\begin{eqnarray}\label{Ex:g=}
&& g^{Is}= \delta^{I\hat{J}}\delta^s_{\hat{J}}\; , \qquad g_J^r= \delta^r_{\hat{J}}\delta_I^{\hat{J}}\; , \qquad
\\ \label{Ex:d=}
&& d^I_{rs}= (0, d^{\hat{I}}_{rs})= \left( 0\; , \; \delta_{(r}^a\delta_{s)}^{\hat{J}} T_{a\, \hat{J}}{}^{\hat{I}}
\right)\; , \qquad d_{Irs}=(d^{\hat{I}}_{rs}\; ,\; 0)= \left( \delta_{(r}^a\delta_{s)}^{\hat{J}} T_{a\,
\hat{J}}{}^{\hat{I}} \; , \; 0 \right)\; , \qquad
\\
\label{frst=} && f_{st}{}^r =  \delta_{s}^b\delta_{t}^cf_{bc}{}^a\delta_{a}^r - \delta_{[s}^a\delta_{t]}^{\hat{J}}
T_{a\,  \hat{J}}{}^{\hat{I}} \delta_{\hat{I}}^r\; , \qquad  \\ \label{Ex:X=} && X_{st}{}^r= -
\delta_{s}^b\delta_{t}^cf_{bc}{}^a\delta_{a}^r + 2 \delta_{[s}^a\delta_{t]}^{\hat{J}} T_{a\,  \hat{J}}{}^{\hat{I}}
\delta_{\hat{I}}^r \, , \qquad X_{rJ}{}^I= \delta_{a}^r\, T_{a\,  \hat{J}}{}^{\hat{I}}
\left(\delta^I_{\hat{I}}\delta_{J}^{\hat{J}}- \delta^{I\hat{I}} \delta_{J\hat{J}}\right) \; . \qquad
\end{eqnarray} Notice that both of Eqs. (\ref{Ex:X=}) give $X_{a\hat{J}}{}^{\hat{I}}=
T_{a\hat{J}}{}^{\hat{I}}$.

Finally, the 3--form fields take values in the $\mathcal R'$ representation only, i.e. \be\label{C3}
C_{3r}=(C_{3\hat I},0)\,. \ee For simplicity, in the further consideration we shall not take into account tensor
fields which are singlets with respect to the non--Abelian symmetries.

The field strengths \eqref{cF:=}-\eqref{cH4:=} take the following form
\begin{eqnarray}\label{cF:=m}
&&{\cal F}^r:\quad \mathcal F^a=  dA^a +{1\over 2} f_{bc}{}^a A^b\wedge A^c \; ,\quad
\mathcal F^{\hat I}=d\mathcal A^{\hat I}+\frac 12\mathcal A^{\hat I}\wedge A^a T_{a\hat J}{}^{\hat I}+B_2^{\hat I} \equiv \mathcal B_2^{\hat I}, \\ \label{cH3:=m} &&{\cal H}_3^I:\qquad
\mathcal H^{\hat I}_3= D\mathcal B_2^{\hat I}\,,\qquad \mathcal H_{3 \hat I}= d \hat B_{2\hat I} -\frac 12 T_{a\hat J}{}^{\hat I} A^a\wedge \hat B_{\hat I} + C_{3\, \hat
I}\equiv \mathcal C _{3\, \hat I} , \\ \label{cH4:=m} &&{\cal H}_{4\hat r}:\qquad {\cal H}_{4\hat I}=D\mathcal C
_{3\, \hat I}\,,
\end{eqnarray}
where the covariant derivative $D$ contains the vector potential $A^a$ only,
$$
D=d+A^aT_a\,  \qquad
$$
or more explicitly
 \bea\label{covariantD}
 &&D\mathcal B_2^{\hat I}= d\mathcal B_2^{\hat I} +\mathcal B_2^{\hat J}\wedge  A^{a}
T_{a\hat{J}}{}^{\hat I} , \nonumber\\
 && D\mathcal C_{3\hat I}= d\mathcal C_{3\hat I} -  \mathcal C_{3\hat J} \wedge A^{a} T_{a\hat{I}}{}^{\hat J} \,.
\eea

Note that the fields $\mathcal A^{\hat I}$ and $\hat B_{2\hat I}$ are of a St\"uckelberg type and thus can be
absorbed, respectively, by $B_2^{\hat I}$ and $C_{3\hat I}$, which is indicated in \eqref{cF:=m} and
\eqref{cH3:=m} by renaming $\mathcal F^{\hat I}\equiv \mathcal B_2^{\hat I} $ and $\mathcal H_{3 \hat I}\equiv
\mathcal C _{3\, \hat I}$. It is these latter fields that transform covariantly under the gauge--group representations generated by $(T_a)_{\hat J}{}^{\hat I}$ and enter the action.

In this case the action \eqref{miniL}-\eqref{L6HH=} reduces to the following form \bea\label{miniex}
S=\int_{\mathcal M_6}\left (-D\hat\phi_{\hat I} \wedge *D\phi^{\hat I}+d^6x \hat\phi_{\hat{I}} (T_a)_{\hat
J}{}^{\hat I}Y^{aij}Y^{\hat J}_{ij} + 2\hat\phi_{\hat I}(T_a)_{\hat J}{}^{\hat I} {\cal B}_2^{\hat J}\wedge *{\cal
F}^a \right)\nonumber\\ -\int_{\mathcal M_6}\left( i_v({\cal H}^{\hat I}_3+ *{\cal H}^{\hat I}_{3}) \wedge {\cal
C}_{3\hat I}\wedge v + i_v({\cal C}_{3\hat I}+ *{\cal C}_{3\hat I}) \wedge {\cal H}^{\hat I}_3\wedge v- {\cal
H}_{3}^{\hat I}\wedge{\cal C}_{3\hat I}\right). \eea The first term in \eqref{miniex} can be rewritten in the
following form \be\label{dual1} i_v({\cal H}_{3}^{\hat I}+ *{\cal H}_{3}^{\hat I}) \wedge {\cal C}_{3\hat I}\wedge
v=i_v({\cal C}_{3\hat I}+ *{\cal C}_{3{\hat I}}) \wedge {\cal H}_{3}^{\hat I}\wedge v+{\cal C}_{3\hat I}\wedge
{\cal H}_{3}^{\hat I}. \ee So the action \eqref{miniex} takes the form
 \bea\label{miniex12} S=\int_{\mathcal
M_6}\left (-D\hat\phi_{\hat I} \wedge *D\phi^{\hat I}+ d^6x \hat\phi_{\hat{I}} (T_a)_{\hat J}{}^{\hat
I}Y^{aij}Y^{\hat J}_{ij} + 2\hat\phi_{\hat I}(T_a)_{\hat J}{}^{\hat I} {\cal B}_2^{\hat J}\wedge *{\cal F}^a
\right)  \nonumber \\ +2\int_{\mathcal M_6}{\cal H}_{3}^{\hat I}\wedge ({\cal C}_{3\hat I} - i_v({\cal C}_{3\hat
I}+ *{\cal C}_{3\hat I})  \wedge v)\; .\qquad \eea Now note that the combination of the ${\cal C}_{3\hat I}$ terms
is anti--self--dual. Indeed, in view of the identity \eqref{v-id} \be\label{C-} C^-_{3\hat I}:= {\cal C}_{3\hat I}
- i_v({\cal C}_{3\hat I}+ *{\cal C}_{3\hat I})  \wedge v=-i_v*{\cal C}_{3\hat I}\wedge v + *(i_v*{\cal C}_{3\hat
I}\wedge v)=-*C^-_{3\hat I}\,. \ee
The generic identity \eqref{v-id} applied to an anti--self dual tensor reads
\be\label{C-id} C^-_{3\hat I}=i_vC^-_{3\hat I}\wedge v - *(i_v{C}^-_{3\hat I}\wedge v). \ee
{}From \eqref{C-} and
\eqref{C-id} it follows that
\be\label{*C3} *{\cal C}_{3\hat I}=-C^-_{3\hat I}+*(\varphi_{2\hat I}\wedge v)\qquad
\Rightarrow \qquad {\cal C}_{3\hat I}=C^-_{3\hat I}+\varphi_{2\hat I}\wedge v, \qquad  \varphi_{2\hat I}=i_v({\cal
C}_{3\hat I}+ *{\cal C}_{3\hat I}) .
\ee
The $\varphi_{2\hat I}\wedge v$  part of ${\cal C}_{3\hat I}$ does not
contribute to the action, so without loss of generality, in \eqref{miniex12} we can replace ${\cal C}_{3\hat I}$
with $C^-_{3\hat I}$ and the action reduces to
 \bea\label{miniex13} S=\int_{\mathcal
M_6}\left (-D\hat\phi_{\hat I} \wedge *D\phi^{\hat I}+ d^6x \hat\phi_{\hat{I}} (T_a)_{\hat J}{}^{\hat
I}Y^{aij}Y^{\hat J}_{ij} + 2\hat\phi_{\hat I}(T_a)_{\hat J}{}^{\hat I} {\cal B}_2^{\hat J}\wedge *{\cal F}^a
\right) +2\int_{\mathcal M_6}{\cal H}_{3}^{\hat I}\wedge {C}^-_{3\hat I} \; .\qquad \eea We see that the auxiliary
one--form field $v(x)$ completely disappears from the action and the anti--self dual field ${C}^-_{3\hat
I}=-*{C}^-_{3\hat I}$ is the Lagrange multiplier which ensures the anti--self--duality of $\mathcal H^{\hat I}_3$.
On the other hand, the variation of this action with respect to ${\cal B}_{2}^{\hat I}$ produces the duality
relation between the field strengths of ${C}^-_{3\hat I}$ and $A^a$ \bea\label{vcB21} D{C}^-_{3\hat I} +  *{\cal
F}^a \, \hat\phi_{\hat J}(T_a)_{\hat I}{}^{\hat J}=0\,. \eea The variation with respect to $A^a$ gives the
equation of motion \be\label{A}
 \left(\hat\phi_{\hat J}\, D* {\cal B}_2^{\hat{I}} + * {\cal B}_2^{\hat{I}}\wedge D\hat\phi_{\hat J}
 -2 {\cal
 B}_2^{\hat{I}}\wedge {C}^-_{3\hat{J}}   + \frac 12  \phi^{\hat I} *D\hat\phi_{\hat J} +\frac 12
 \phi_{\hat J}
 *D\hat \phi^{\hat I}\right) T_{a\,  \hat{I}}{}^{\hat{J}} =0\; .
\ee Yang--Mills type equations for the vector gauge fields can be obtained as a self--consistency condition for
eq. (\ref{vcB21}), \be\label{ddA}
 \left(\hat\phi_{\hat J}\, D* {\cal F}^{a} + * {\cal F}^{a}\wedge D\hat \phi_{\hat J} -{1}/{2} {\cal
 F}^{a} \wedge {
 C}^-_{3\hat{J}}  \right) T_{a\,  \hat{I}}{}^{\hat{J}} =0\; .
\ee Finally, the scalar field equations are \be\label{phiEq}
 D*D\hat \phi_{\hat I}=0\,,\qquad D*D\phi^{\hat I}=T_{a\hat J}{}^{\hat
I}\left(d^6x \, Y^{aij}Y^{\hat J}_{ij} + 2 {\cal B}_2^{\hat{J}}  \wedge  * {\cal F}^{a}\right)\;, \ee \be\label{Y}
\hat\phi_{\hat{I}} (T_a)_{\hat J}{}^{\hat I}Y^{aij}=0\,,\qquad \hat\phi_{\hat{I}} (T_a)_{\hat J}{}^{\hat I}Y^{\hat
J}_{ij}=0\,. \ee

\section{Conclusion}
\label{sec:conclusions}

We have constructed the duality--symmetric actions for a large class of six--dimensional models describing hierarchies of non--Abelian scalar, vector and tensor fields related to each other by (self-)duality equations that follow from these actions. This class includes the bosonic sectors of the $6d$ (1,0) superconformal models of interacting non--Abelian vector, tensor and hypermultiplets constructed in \cite{Samtleben:2011fj,Samtleben:2012mi,Samtleben:2012fb}. The supersymmetrization of the actions of this paper by the inclusion of fermionic sectors will be considered elsewhere. A generic feature of the supersymmetric manifestly duality--invariant actions is that the off--shell supersymmetry transformations of fermionic fields get augmented by terms which vanish when the bosonic fields satisfy the (self-)duality conditions (see \emph{e.g.} \cite{Schwarz:1993vs,Pasti:1995ii}).

We have first obtained the action~(\ref{miniL}) that gives rise to non-Abelian self-duality equations for the tensor fields. In the second step, we have extended this action to the action (\ref{Lext}) that also yields the non-Abelian first-order duality equations between vector and three-form tensor gauge potentials. Continuing this line of thought, a natural next step in the construction would be the extension of (\ref{Lext}) to an action that also yields the first-order duality equations between scalar and four-form tensor gauge potentials. This would correspond to a truly democratic formulation of the six-dimensional models, in which {\em all} $p$-forms enter on equal footing with the forms of different degree interlocked by the non-Abelian structure of the tensor hierarchy. This final extension to include the duality equations for the scalar fields will proceed straightforwardly along the pattern put forward in Section~\ref{sec:action}. On the technical side it will require to extend the six-dimensional tensor hierarchy of Section~\ref{sec:hierarchy} by the inclusion of five-form gauge potentials, c.f.~\cite{deWit:2008gc}.

In connection with the issues of the (2,0) superconformal theory of multiple $M5$--branes,
further study is required for understanding whether in some of these (1,0) supersymmetric models the redundant degrees of freedom associated with propagating vector fields can be removed and (1,0) supersymmetry can be enhanced to (2,0). Another important issue to be resolved is the presence (in general) of ghosts in the action due to the non-positive definiteness of the metric $\eta_{IJ}$ (see e.g. eq. \eqref{Lphi}).
Clearly, it would also be of interest to study the relation of these systems to other proposals of non--Abelian $6d$ chiral
tensor models and, by dimensional reduction, to $5d$  and $4d$ super--Yang--Mills theories.

\section*{Acknowledgements}
The authors are grateful to David Berman, Chong-Sun Chu, Neil Lambert, Kimyeong Lee, George Papadopoulos, Paolo Pasti, Christian Saemann, Igor Samsonov, Ergin Sezgin, Douglas Smith, Thomas Strobl, Mario Tonin, Robert Wimmer and Martin Wolf for interest to this work and useful discussions.
This work was partially supported by the Padova University Research Grant CPDA119349. D.S. was also supported in part by the
MIUR-PRIN contract 2009-KHZKRX. I.B. was supported in part by the research grant FPA2012-35043-C02-01 from the MICINN (presently MEC) of Spain and by the Basque Government Research Group Grant ITT559-10.
H.S. is grateful to INFN and the University of Padova for kind hospitality while this project was started.
I.B. and D.S. acknowledge kind hospitality and support extended to them at the Galileo Galilei Institute Workshop program ``Higher spins, strings and dualities" (Florence, March 13 - May 10, 2013) during the final stage of this project.

\begin{appendix}

\section{Algebraic constraints on the constant tensors}
\label{app:con}

The algebraic consistency conditions for the tensors $f_{st}{}^r$, $d^I_{rs}$, $g^{Ir}$, $k^\alpha_r$ defining the
six-dimensional tensor hierarchy are given by  \bea d_{I\,r(u}d^I_{vs)} &=& 0 \;,\nonumber\\[.4ex] {}
\left(d^J_{r(u}\, d^I_{v)s} -d^J_{uv}\,d^I_{rs} + d_{K\,rs} d^K_{uv}\, \eta^{IJ}\right)\tg_J^s &=& f_{r(u}{}^s
d^I_{v)s} \;,\nonumber\\[.4ex] 3f_{[pq}{}^u f_{r]u}{}^s -\tg^s_I\, d^{I}_{u[p} f_{qr]}{}^u &=& 0
\;,\nonumber\\
[.4ex] {} X_{rs}{}^t\equiv d^I_{rs}\,g_I^t-f_{rs}{}^t&=&-k_r^\alpha c_{\alpha\,s}^t\nonumber\\
[.4ex] {}X_{r\,IJ} \equiv 4 g_{[I}^sd_{J]\,rs}   &=& 2\,k_r^\alpha\,c_{\alpha\,IJ}\nonumber\\
[.4ex] {} f_{rs}{}^t \tg_I^r -  d^J_{rs}\,\tg_J^t \tg_I^r &=& 0 \;,\nonumber\\[.4ex] g_K^r
\tg_{[I}^{s}d^{\vphantom{s}}_{J]sr} &=& 0 \;,\nonumber\\[.4ex] \tg_I^r \tg^{Is} &=& 0 \;,
\nonumber\\[.4ex] k^\alpha_r g^{Ir} &=& 0\,.
\label{conaction}
\eea
In particular, the third equation shows that the violation of the Jacobi identities of the `structure constants'
$f_{rs}{}^t$ is related to the St\"uckelberg coupling $g_I^r$\,.
The general structure of solutions to these constraints has been analyzed in~\cite{Samtleben:2012mi}.

\end{appendix}

\if{}
\bibliography{references}
\end{document}
\fi

\end{document}